\documentclass[doublecol]{epl2}

\usepackage{epsfig}

\usepackage{graphicx}

\def\be{\begin{equation}}
\def\ee{\end{equation}}

\title{Phase coexistence in polydisperse mixture of hard-sphere colloidal
and flexible chain particles}
\shorttitle{Polydisperse polymer-colloidal mixture}
\author{Yurij V. Kalyuzhnyi\inst{1} \and Peter T. Cummings\inst{2}}
\shortauthor{Yu.V. Kalyuzhnyi {\it et al.}}
\institute{
\inst{1} Institute for Condensed Matter Physics, Svientsitskoho 1, 79011
  Lviv, Ukraine \\
\inst{2}Department of Chemical Engineering,
  Vanderbilt University, Nashville, TN 37235-1604, USA and Chemical
  Sciences Division, Oak Ridge National Laboratory, Oak Ridge, TN
  27831-6110, USA}

\abstract{
A theoretical scheme for the calculation of the full phase diagram
(including cloud and shadow curves, binodals and distribution functions
of the coexisting phases) for colloid-polymer mixtures with polymer
chain length polydispersity and hard-sphere colloidal and polymeric
monomer sizes polydispersity is proposed. The scheme combines
thermodynamic perturbation theory for associating fluids
and recently developed method used to determine the phase diagram
of polydisperse spherical shape colloidal fluids
(L.Bellier-Castella {\it et al.},
{\it J.Chem.Phys.} {\bf 113}, 8337(2000)). By way of illustration we
present and discuss the full phase diagram for the mixture with
polydispersity in the size of the hard-sphere colloidal particles.}

\begin{document}

\maketitle

\section{Introduction}

Since the pioneering studies of Asakura and Oosawa \cite{AO1,AO2} and
Vrij \cite{vrij} substantial amount of efforts have been focused on the
development of the theoretical methods describing
the phase behavior of the athermal colloid-polymer mixtures (see Refs.
\cite{tuiner,fuchs,paricaud1} and references therein).
Usually in the vast majority of the theories developed so far interaction between
polymers is either ignored or treated using approximations different from
those assumed to describe polymer-colloid and colloid-colloid interactions.
This feature imposes certain restrictions on the possibilities of the theory,
for example the theories, which ignore polymer-polymer interactions are
restricted to the mixtures of short polymers and large colloidal particles
(ideal polymer limit).
Recently an attempt to describe the phase behavior of the colloid-polymer
mixture with both components treated on an equal footing using polymer reference
interaction site model integral-equation approach \cite{zukowski} and
thermodynamic perturbation theory (TPT) for associating fluids \cite{paricaud1}
was made. In these studies polymer molecules were modeled as a flexible chains
of tangentially bonded hard-sphere monomers and colloidal particles were
represented as a hard spheres.
Later TPT approach was extended \cite{paricaud2} to account for the
chain length polydispersity effects. In this study we propose further
extension of the TPT to account for polydispersity in the polymer chain length and
in the hard-sphere sizes of both colloidal particles and polymer monomers. This
extension is based on the previously developed method \cite{xu,kal1}, which allows
us to determine the full phase diagram (including binodals and cloud and shadow
curves) and to discuss fractionation effects on the level of the
distribution functions of the two daughter phases.

\section{The model}

We consider polydisperse mixture of hard-sphere flexible chain particles
represented by $m$ tangentially bonded hard spheres of diameter $\sigma$.
In the following we will distinguish between $p-$type of the particles
(polymers) and $c-$type of the particles (colloids). Thus the species of
each particle is characterized by the set of three variables $(a,m,\sigma)$
with $a$ denoting the type of the particle (either $p$ or $c$),
$m=1,2,...,\infty$ and $0\le \sigma <\infty$. The number density of the $a-$type
of the particles is $\rho_a$ and the overall number density $\rho$ is $\rho=\rho_p+\rho_c$.
The species variables $a$, $m$ and $\sigma$ are
distributed according to the distribution function $F_a(m,\sigma)$, which
is positive and satisfies the following normalizing conditions
\be
\sum_a\sum_m\int d\sigma\;F_a(m,\sigma)=1.
\label{norm0}
\ee
Further we put $F_a(m,\sigma)=\alpha_af_a(m,\sigma)$, where $\alpha_a$ denote
the fraction of the $a-$type of the particles
\be
\alpha_a=\sum_m\int d\sigma\;F_a(m,\sigma);
\label{alpha}
\ee
obviously, $\alpha_p+\alpha_c=1$ and partial distribution functions
$f_a(m,\sigma)$ are normalized.

\section{Thermodynamical properties}

Thermodynamical properties of the model at hand are calculated using TPT
of Wertheim \cite{wtd,gubbins}. According to TPT Helmholtz free energy
of the system $A$ is represented as a sum of three terms
$A=A_{id}+A_{hs}+ A_{ch}$,
where $A_{id}$ is the ideal gas term
\be
{A_{id}\over VkT}=\sum_a\rho_a\sum_m\int d\sigma\;F_a(m,\sigma)
\left\{\ln\left[\rho_aF_a(m,\sigma)\right]-1\right\},
\label{Aid}
\ee
$A_{hs}$ is the hard-sphere term
\be
{ A_{hs}\over VkT}={6\over \pi}\left[\left({\zeta_2^3\over \zeta_3^2}-\zeta_0\right)
\ln \Delta+{3\zeta_1\zeta_2\over \Delta}+
{\zeta_2^3\over \zeta_3\Delta^2}\right],
\label{Ahs}
\ee
and $ A_{ch}$ is the term describing formation of the chains
\be
{ A_{ch}\over VkT}=\sum_a\rho_a\sum_m \left(1-m\right)   \int d\sigma\;F_a(m,\sigma)
\ln g_{aa}^{(hs)}(\sigma).
\label{Ach}
\ee
Here $V$ is the system volume, $k$ is the Boltzmann constant, T is the temperature,
$\zeta_0,\zeta_1,\zeta_2,\zeta_3$ are the distribution function moments
\be
\zeta_n={\pi\over 6}\rho\sum_a\sum_mm\int d\sigma\;F_a(m,\sigma)\sigma^n,
\hspace{5mm} n=0,1,2,3;
\label{zeta}
\ee
$\Delta=1-\zeta_3$, $g^{(hs)}_{aa}(\sigma)$ is the hard-sphere contact value
\be
g_{aa}^{(hs)}(\sigma)={1\over \Delta}\left(1+{3\over 2}\sigma
{\zeta_2\over \Delta}
+{1\over 2}\sigma^2{\zeta_2^2\over \Delta^2}\right).
\label{gaa}
\ee
In (\ref{Ahs}) and (\ref{gaa}) Boublik's
\cite{boublik} and Mansoori's {\it et al.} \cite{mansoori}
expressions for $A_{hs}$ and $g_{aa}^{(hs)}(\sigma)$ have been utilized.
All the rest of thermodynamical properties can be obtained from Helmholtz free energy
using the standard thermodynamical relations. Taking the volume derivative of
the free energy we get the following expression for the pressure:
$\beta P=\rho+\beta P_{hs}+\beta P_{ch}$,
where
\be
\beta P_{hs}={6\over \pi\Delta}\left\{\zeta_0\zeta_3+
{\zeta_2\over \Delta}\left[3\zeta_1+
{\zeta_2^2\over \Delta}\left(2+\Delta\right)\right]\right\},
\label{Phs}
\ee
\be
\beta P_{ch}={1\over \Delta}\left(\zeta_3\Omega+\zeta_2\Psi\right),
\label{PhsPch}
\ee
\be
\Omega=\rho\sum_a\sum_m\left(1-m\right)\int d\sigma\;F_a(m,\sigma)\Omega(\sigma),
\hspace{3mm}
\label{Om}
\ee
\be
\Psi=\rho\sum_a\sum_m\left(1-m\right)\int d\sigma\;F_a(m,\sigma)\Psi(\sigma),
\label{Psi1}
\ee
\be
\Omega(\sigma)=
{\Delta^2+3\sigma\zeta_2\left(\Delta+{1\over 2}\sigma\zeta_2\right)
\over \Delta^2+{3\over 2}\sigma\zeta_2\left(\Delta+{1\over 3}\sigma\zeta_2\right)},
\label{Om2}
\ee
\be
\Psi(\sigma)=
{{3\over 2}\sigma\Delta\left(\Delta+{2\over 3}\sigma\zeta_2\right)
\over \Delta^2+{3\over 2}\sigma\zeta_2\left(\Delta+{1\over 3}\sigma\zeta_2\right)}.
\label{Psi2}
\ee
Chemical potential is obtained as the functional derivative of the
free energy density $F/V$ with respect to the function
$\rho_a(m,\sigma)=\rho_aF_a(m,\sigma)$:
\be
\mu_a(m,\sigma)=\ln \left[\rho_aF(m,\sigma)\right]
+\mu_{a,hs}(m,\sigma)+\mu_{a,ch}(m,\sigma),
\label{mu}
\ee
where
$$
\beta\mu_{a,hs}(m,\sigma) =
m\left[\sigma^2{\zeta^2_2\over \zeta^2_3}
\left(3-2\sigma{\zeta_2\over \zeta_3}\right)-1\right]\ln\Delta
$$
$$
+{\pi m\over 2\Delta}\sigma\left\{\sigma^2
\left[{2\over \pi}\zeta_0-
\frac{\zeta^3_2}{\zeta^2_3}
\frac{\left(1+\Delta\right)}{3\Delta}+{\pi\over 3\Delta}
\zeta_2\left(\frac{1}{2}\zeta_1+
\frac{1}{3} \frac{\zeta^2_2}{\zeta_3\Delta}\right)\right]
\right.
$$
\be
\left.
+\zeta_2
\left(1+\sigma\frac{\zeta_2}{\zeta_3\Delta}\right)+\sigma\zeta_1
\right\},
\label{mhs}
\end{equation}
\be
\beta\mu_{a,ch}(m,\sigma)=\left(1-m\right)\ln g_{aa}(\sigma)+
{\pi\over 6}{m\sigma^2\over \Delta}\left(\Omega\sigma+\Psi\right).
\label{mch}
\ee
One can easily see that thermodynamical properties of the model at hand
are defined by the set of a finite number of the distribution function moments,
i.e. four regular moments $\zeta_0,\zeta_1,\zeta_2,\zeta_3$ and two generalized
moments $(\Omega,\;\Psi)$. Thus polydisperse mixture of the chain particles treated
within TPT belong to the class of truncatable free energy models \cite{sollich}.

\begin{figure}
\centering
\epsfig{file=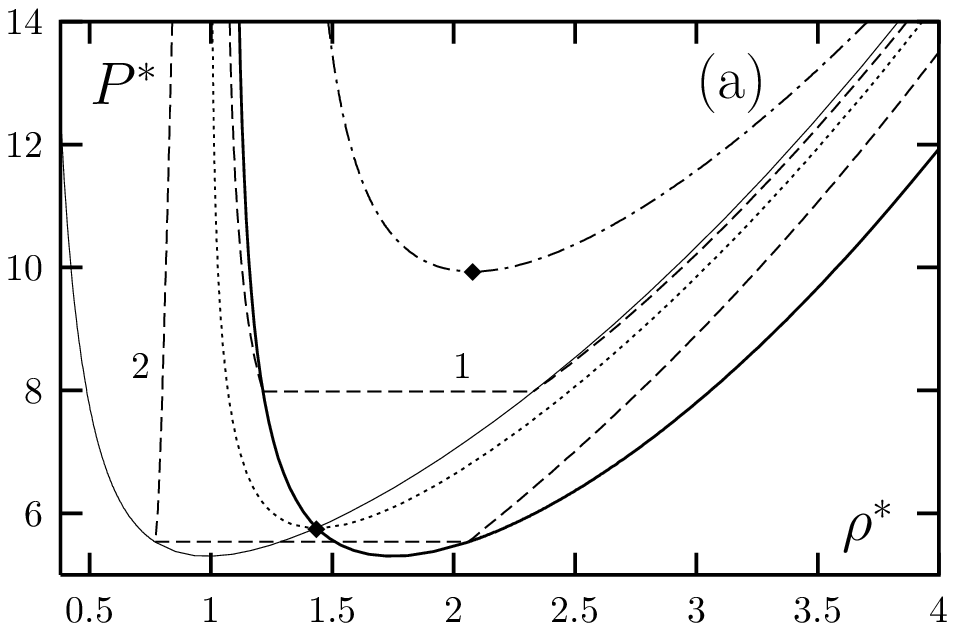,width=0.485\textwidth} \hfill
\epsfig{file=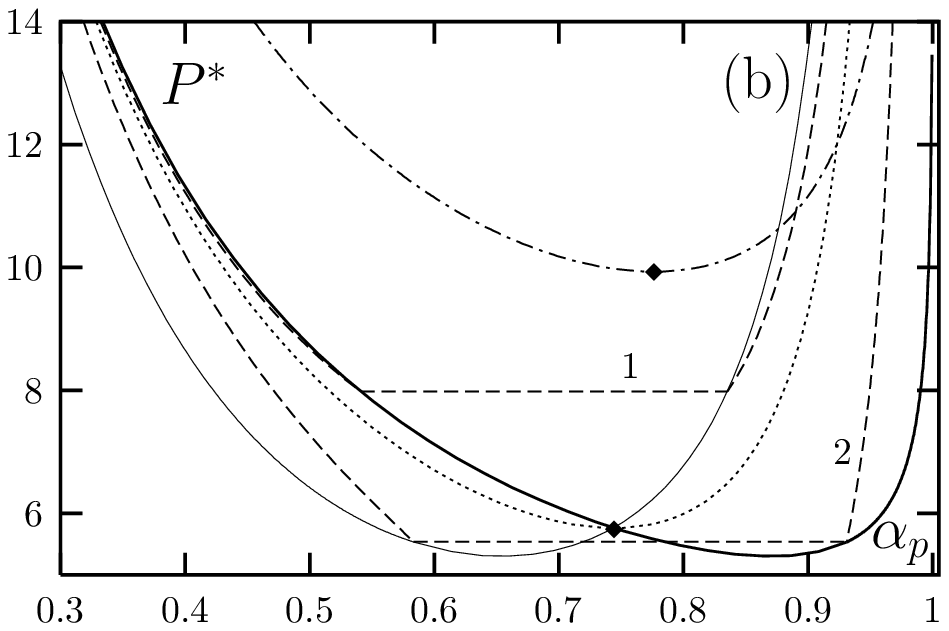,width=0.485\textwidth}
\caption{Phase diagram in: (a) $P^*$ vs $\rho^*$ and (b) $P^*$ vs $\alpha_p$
coordinate frames. Cloud and shadow curves are represented by thick and thin
solid lines, respectively, binodal curves with: (1) $\alpha_p^{(0)}=0.541$ and
(2) $\alpha_p^{(0)}=0.931$ by the broken lines and with
$\alpha_p^{(0)}=\alpha^{(0)}_{p,cr}=0.744$ by the dotted line.
The dashed-dotted line denotes the binodal curve
for the bidisperse version of the model and critical points are indicated by
the diamonds.}
\label{f1}
\end{figure}

\section{Phase equilibrium conditions}

We assume that at a certain density $\rho^{(0)}$ and composition $F_a^{(0)}(m,\sigma)$
the system separates into two phases with the densities $\rho^{(1)}$ and $\rho^{(2)}$,
 and
compositions $F_a^{(1)}(m,\sigma)$ and $F_a^{(2)}(m,\sigma)$. Hereafter the upper
index $(0)$ refer to the parent phase and the upper indices $(1)$ and $(2)$
refer to the daughter phases. At equilibrium these quantities take the values, which
follows from the phase equilibrium conditions, i.e.: (i) conservation of the
total volume of the system, (ii) conservation of the total number of the particles of
each species, (iii) equality of the chemical potentials of particles of the same
species in the coexisting phases, (iv) equality of the pressure in the coexisting
phases. These conditions finally lead to the following set of relations
\cite{xu,kal1}:
$$
F^{(\alpha)}_a(m,\sigma)=F^{(0)}_a\left(m,\sigma\right)
\hspace{50mm}
$$
\be
\hspace{12mm}
\times Q_a^{(\alpha)}\left(m,\sigma;\rho^{(0)},
\rho^{(1)},\rho^{(2)};\left[F_p^{(\alpha)},F_c^{(\alpha)}\right]\right),
\label{FQ}
\ee
\be
P^{(1)}\left(\rho^{(1)};\left[F_p^{(1)},F_c^{(1)}\right]\right)=
P^{(2)}\left(\rho^{(2)};\left[F_p^{(2)},F_c^{(2)}\right]\right),
\label{P1P2}
\ee
\be
\sum_a\sum_m\int d\sigma\;F^{(\alpha)}_a(m,\sigma)=1,
\hspace{3mm} {\rm for}\;\alpha=1\;{\rm or} \;\alpha=2,
\label{norm12}
\ee
where
$$
\rho^{(\alpha)}Q_a^{(\alpha)}\left(m,\sigma;\rho^{(0)},\rho^{(1)},\rho^{(2)};
\left[F_p^{(\alpha)},F_c^{(\alpha)}\right]\right)
$$
\be
=
{\rho^{(0)}\left(\rho^{(2)}-\rho^{(1)}\right)\left[1-\delta_{1\alpha}+
\delta_{1\alpha}\exp \left(\beta\Delta{\tilde \mu}_a\right)\right]\over
\rho^{(0)}-\rho^{(1)}-\left(\rho^{(0)}-\rho^{(2)}\right)
\exp \left(\beta\Delta{\tilde \mu}_a\right)},
\label{Q}
\ee
\be
\Delta{\tilde \mu}_a
={\tilde \mu_a}^{(2)}\left(m,\sigma,\rho^{(2)};\left[F_a^{(2)}\right]\right)
-{\tilde \mu_a}^{(1)}\left(m,\sigma,\rho^{(1)};\left[F_a^{(1)}\right]\right),
\label{Dmu}
\ee
${\tilde \mu}_a^{(\alpha)}$ is the excess (over the ideal gas) chemical potential
of the particle $(a,m,\sigma)$ in the phase $\alpha$ and $\left[\ldots\right]$ denote
functional dependence. The relation between $F^{(0)}_a(m,\sigma)$ and daughter phase
distribution function $F^{(\alpha)}_a(m,\sigma)$, i.e., Eq. (\ref{FQ}), follows from
the phase equilibrium conditions (i)-(iii).

Relations (\ref{FQ})-(\ref{norm12}) represent a closed set of equations to be solved
for the unknowns $\rho^{(\alpha)}$ and $F^{(\alpha)}_a(m,\sigma)$; this set have to
be solved for every value of the species variables $a$, $m$ and $\sigma$. However,
since thermodynamical properties of the model at hand are defined by the finite number
of the moments we can map this set of equations onto a closed set of fourteen
algebraic equations for $\rho^{(\alpha)}$ and twelve moments $\zeta_n^{(\alpha)}$
$(n=0,1,2,3)$, $\Omega^{(\alpha)}$ and $\Psi^{(\alpha)}$, where $\alpha=1,2$.
We have
$$
\zeta_n^{(\alpha)}=\rho^{(\alpha)}\sum_a\sum_mm\int d\sigma\;\sigma^nF^{(0)}_a
\left(m,\sigma\right)
\hspace{22mm}
$$
\be
\hspace{21mm}
\times Q_a^{(\alpha)}\left(m,\sigma,\rho^{(0)};
\left\{X^{(1)}\right\},\left\{X^{(2)}\right\}\right),
\label{m1}
\ee
$$
\Omega^{(\alpha)}=\rho^{(\alpha)}\sum_a\sum_m\left(1-m\right)\int d\sigma\;
\Omega\left(\sigma\right)F^{(0)}_a
\left(m,\sigma\right)
\hspace{8mm}
$$
\be
\hspace{21mm}
\times Q_a^{(\alpha)}\left(m,\sigma,\rho^{(0)};
\left\{X^{(1)}\right\},\left\{X^{(2)}\right\}\right),
\label{m2}
\ee
$$
\Psi^{(\alpha)}=\rho^{(\alpha)}\sum_a\sum_m\left(1-m\right)\int d\sigma\;
\Psi\left(\sigma\right)F^{(0)}_a
\left(m,\sigma\right)
\hspace{8mm}
$$
\be
\hspace{21mm}
\times Q_a^{(\alpha)}\left(m,\sigma,\rho^{(0)};
\left\{X^{(1)}\right\},\left\{X^{(2)}\right\}\right),
\label{m3}
\ee
where $\left\{X^{(\alpha)}\right\}$ represent unknowns of the problem, i.e.
$$
\left\{X^{(\alpha)}\right\}=\left\{\rho^{(\alpha)},\zeta_n^{\alpha},
\Omega^{(\alpha)},\Psi^{(\alpha)}\right\},\;n=0,1,2,3;\;\alpha=1,2.
$$
The remaining two equations follows from the equality of the pressure in coexisting
phases (\ref{P1P2}),
\be
P^{(1)}\left(\rho^{(1)};\left\{X^{(1)}\right\}\right)=
P^{(2)}\left(\rho^{(2)};\left\{X^{(2)}\right\}\right),
\label{P1P2X}
\ee
and from the normalizing condition (\ref{norm12}) for either phase
$\alpha=1$ or $\alpha=2$,
$$
\sum_a\sum_m\int d\sigma\;F^{(0)}_a
\left(m,\sigma\right)
\hspace{45mm}
$$
\be
\times Q_a^{(\alpha)}\left(m,\sigma,\rho^{(0)};
\left\{X^{(1)}\right\},\left\{X^{(2)}\right\}\right)=1.
\label{norm}
\ee
Solution of the set of equations (\ref{m1})-(\ref{norm}) for a given density
$\rho^{(0)}$ and distribution function $F_a^{(0)}(m,\sigma)$ of the parent phase
gives the densities $\rho^{(\alpha)}$ and distribution functions
$F_a^{(\alpha)}(m,\sigma)$ of the two coexisting daughter phases. The coexisting
densities at different densities of the parent phase $\rho^{(0)}$ defined binodals,
which are terminated when a density of one of the phases is equal to the parent phase
density $\rho^{(0)}$. For the different values of the colloidal (and polymer)
particles fraction $\alpha_c^{(0)}$ $(\alpha^{(0)}_p=1-\alpha^{(0)}_c)$ in the
parent phase these termination points form the cloud and shadow coexisting
curves. These curves intersect at the critical point, which is characterized by the
critical density $\rho_{cr}=\rho^{(1)}=\rho^{(2)}=\rho^{(0)}$ and critical
colloidal (and polymeric) composition
$\alpha_{c,cr}=\alpha^{(1)}_c=\alpha^{(2)}_c=\alpha^{(0)}_c$
$(\alpha_{p,cr}=1-\alpha_{c,cr})$.
The cloud-shadow curves can be obtained as a special solution of the general
coexisting problem, when the properties of one phase are equal to the properties
of the parent phase: assuming that the phase $\alpha=2$ is the cloud phase, i.e.
$\rho^{(2)}=\rho^{(0)}$, and following the above scheme we will end up with the
same set of equations (\ref{m1})-(\ref{norm}), but with $\rho^{(2)}$ and
$F_a^{(2)}(m,\sigma)$ substituted by $\rho^{(0)}$ and $F_a^{(0)}(m,\sigma)$,
respectively.

\begin{figure}
\centering
\epsfig{file=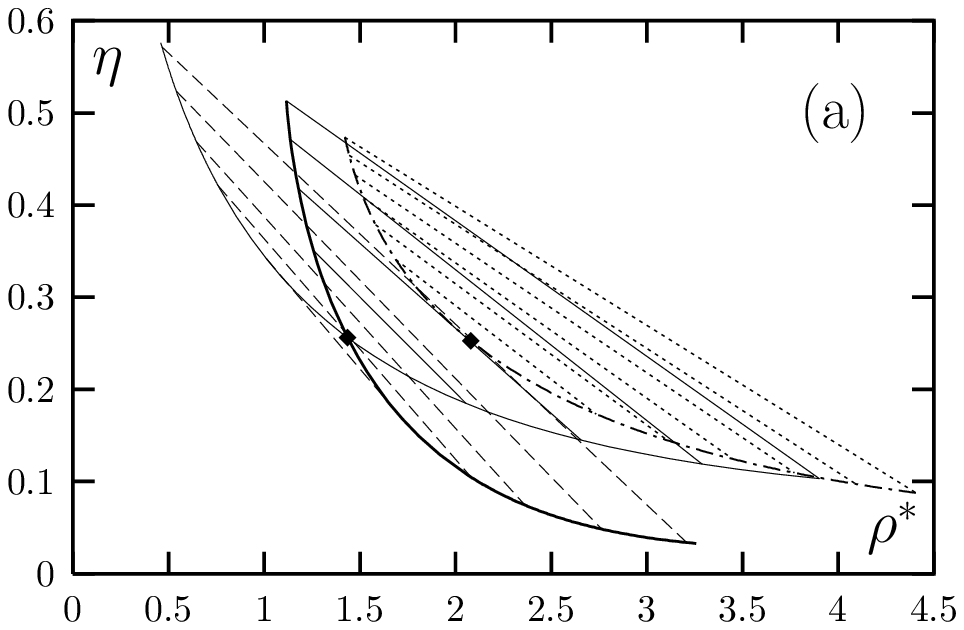,width=0.485\textwidth} \hfill
\epsfig{file=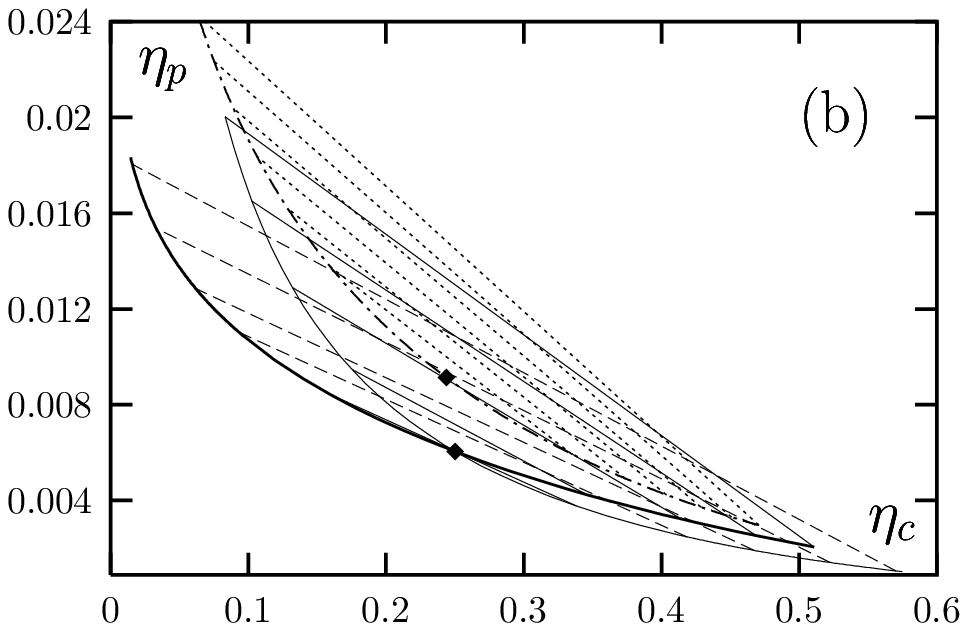,width=0.485\textwidth}
\caption{The same as in Figure 1 in: (a) $\eta$ vs $\rho^*$ and (b)
$\eta_p$ vs $\eta_c$ coordinate frames. Thin solid tie lines and broken tie
lines connect the points on the cloud and shadow curves and dotted tie lines
connect the points on the binodal curve for the bidisperse version of the model.}
\label{f2}
\end{figure}

\section{Results and discussion}

To illustrate the scheme developed in the previous sections we present here
numerical results for the phase behavior of the polydisperse mixture of colloidal
and polymeric particles with poydispersity in the size of the colloidal particles
only. Colloids are represented by the polydisperse mixture of hard spheres and
polymers are modeled by the hard-sphere flexible chains with fixed chain length
$m_0$ and hard-sphere size of the monomeric units $\sigma_p$. Thus for the
distribution function of the parent phase $F_a^{(0)}(m,\sigma)$ we have
\be
F_p^{(0)}(m,\sigma)=\alpha_p^{(0)}\delta\left(\sigma-\sigma_p\right)\delta_{m,m_0},
\hspace{8mm}
\label{Fp}
\ee
\be
F_c^{(0)}(m,\sigma)=\alpha_c^{(0)}f_c^{(0)}(\sigma)\delta_{m,1},
\label{FpFc}
\ee
where for the colloidal diameter distribution $f_c^{(0)}(\sigma)$ we have chosen
beta-distribution, given by
$$
f_c^{(0)}(\sigma)=
B^{-1}(\gamma,\nu)\left({\sigma\over \sigma_{m}}\right)^{\gamma-1}
\left(1-{\sigma\over \sigma_{m}}\right)^{\nu-1}
\theta\left(\sigma_{m}-\sigma\right),
$$
where $\theta(x)$ is the Heaviside step function, $B(\gamma,\nu)$ is the Beta
function \cite{kal1} and $\gamma$ and $\nu$ are related to the first
$\langle\sigma\rangle^{(0)}_c$ and second $\langle\sigma^2\rangle^{(0)}_c$ moments
by
$$
\gamma={\sigma_{m}-\langle\sigma\rangle^{(0)}_c\left(1+D_\sigma^{(0)}\right)\over
D_\sigma^{(0)}\sigma_{m}},
\hspace{3mm}
\nu=\left({\sigma_{m}-\langle\sigma\rangle_c^{(0)}\over
\langle\sigma\rangle^{(0)}_c}\right)\gamma
$$
with
\be
\langle\sigma^n\rangle_c^{(\alpha)}=\int d\sigma\;\sigma^nf_c^{(\alpha)}(\sigma),
\hspace{1mm}
D_\sigma^{(\alpha)}=\langle\sigma^2\rangle_c^{(\alpha)}/
\left(\langle\sigma\rangle_c^{(\alpha)}\right)^2-1.
\label{D}
\ee
Calculations were carried out for the model parameters chosen to be:
$m_0=50$, $\sigma_p/\langle\sigma\rangle_c^{(0)}=0.06$,
$D_\sigma=0.1$ and $\sigma_{m}/\langle\sigma\rangle_c^{(0)}=2$.

\begin{figure}
\centering
\epsfig{file=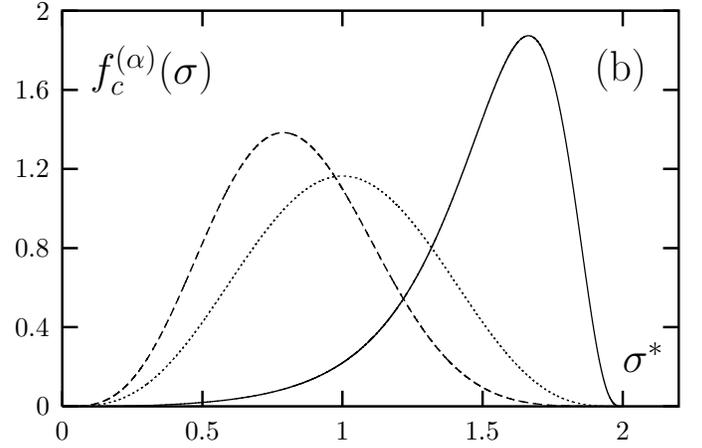,width=0.485\textwidth}
\caption{Distribution functions of the parent phase (dotted line), low density
phase (solid line) and high density phase (dashed line) on the shadow curve at
$P^*=8$. Here $\sigma^*=\sigma/\langle\sigma\rangle^{(0)}_c$}
\label{f4}
\end{figure}

At sufficiently low density of the parent phase $\rho^{(0)}$ the mixture is
stable as a single phase. As $\rho^{(0)}$ is increased the system will phase
separate into two phases: low density phase and high density phase.
In Figures 1 and 2 we show the phase diagram in a different coordinate frames,
i.e. $P^*$ vs $\rho^*$ (Fig. 1a), $P^*$ vs $\alpha_p$ (Fig. 1b),
$\eta$ vs $\rho^*$ (Fig. 2a) and $\eta_p$ vs $\eta_c$ (Fig. 2b)
, where $P^*=\beta P(\langle\sigma\rangle^{(0)}_c)^3$,
$\rho^*=\rho(\langle\sigma\rangle_c^{(0)})^3$,
$\eta$ and $\eta_a$ are the overall packing fraction and packing fraction of
the $a-$type of the particles, respectively. The phase diagram shown in
Figure 1 includes the cloud and shadow curves, along with three
binodals for three selected parent phase fractions
$\alpha^{(0)}_p=0.541,\;0.744,\;0.931$, one of them
being the critical fraction $\alpha^{(0)}_{p,cr}=0.744$.
The two branches of the binodals for the two selected $\alpha_p^{(0)}$ values
($\alpha^{(0)}_p=0.541,\;0.931$) terminate on the cloud and shadow curve, where
the $\rho$ values of the end points of the respective binodals on the
cloud curve are equal to the $\rho^{(0)}$ values of the binodals in these points.
For the critical value of the parent phase fraction $\alpha_{p,cr}^{(0)}=0.744$
corresponding binodals meet in the critical point.
In Figure 2 we include only the cloud and shadow curves.
For the reference in both figures we also
display the binodal curve for the bidisperse version of the model
$(D^{(0)}_\sigma=0)$.
Comparison between the bidisperse and polydisperse versions of the
colloid-polymer mixture shows that polydispersity
extends the region of the phase instability, shifting the critical
point to the lower values of the pressure and density
($P^*_{cr,bid}=9.93,\;P^*_{cr,polyd}=5.74$,
$\rho^*_{cr,bid}=2.08,\;\rho^*_{cr,polyd}=1.43$, Fig. 1)
and to slightly lower values of the polymer fraction
($\alpha^{(bid)}_{p,cr}=0.776,\;\alpha^{(polyd)}_{p,cr}=0.744$, Fig. 1b) and
polymer packing fraction
($\eta^{(bid)}_{p,cr}=0.009,\;\eta^{(polyd)}_{p,cr}=0.006$, Fig. 2b).
Corresponding relative changes of the colloidal ($\eta_{c,cr}$) and
overall ($\eta_{cr}$) critical packing fraction are negligible.

\begin{figure}
\centering
\epsfig{file=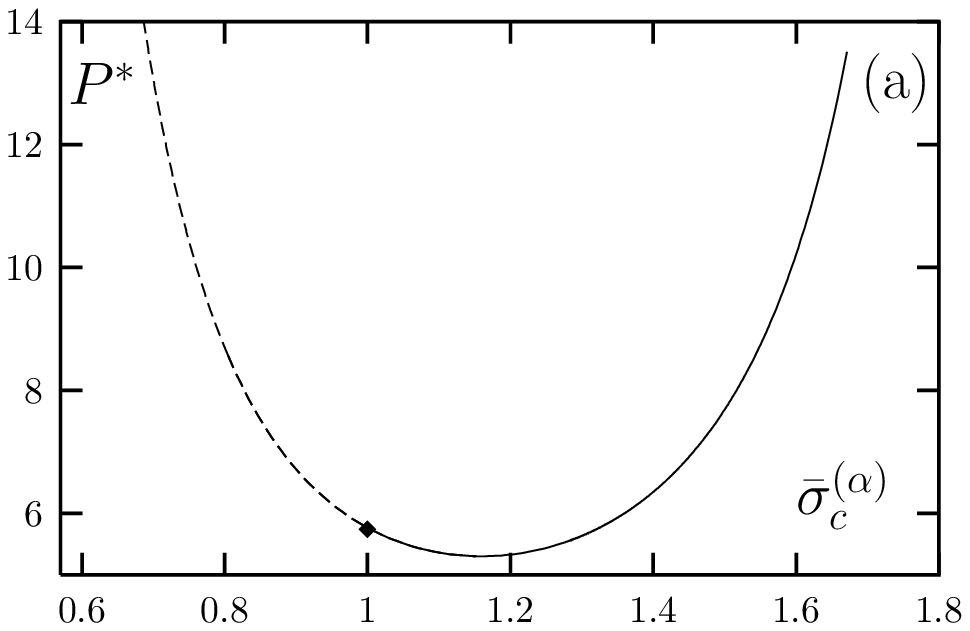,width=0.485\textwidth} \hfill
\epsfig{file=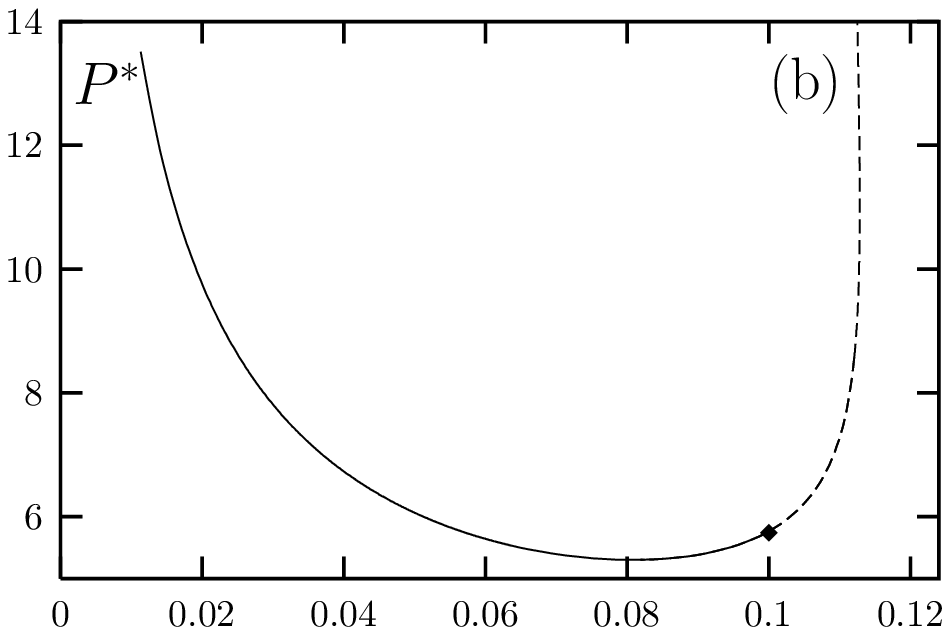,width=0.485\textwidth}
\caption{Average hard-sphere size ${\tilde \sigma}_c^{(\alpha)}
=\langle\sigma\rangle^{(\alpha)}_c/\langle\sigma\rangle^{(0)}_c$ (a)
and distribution function width $D_\sigma^{(\alpha)}$ (b) of the colloidal
particles along the shadow curve. Solid lines represent low density phase and
dashed lines denote high density phase. Critical points are indicated by
the diamonds.}
\label{f3}
\end{figure}

According to Figure 2 packing fraction of the low density phase is higher than
that of the high density phase; thus the density of the liquid-like phase is
lower than that of the gas-like phase. This is not surprising as the fraction of
the colloidal particles in the low density phase is larger than in the high
density phase.
In Figure 3 we show daughter phase distribution functions
$f^{(\alpha)}_c(\sigma)$ on the shadow curve at $P^*=8$
together with the parent phase distribution function $f_c^{(0)}(\sigma)$.
More detailed information about the composition of the coexisting phases
can be extracted from the analysis of the distribution functions of the two
daughter phases in terms of their first two moments,
$\langle\sigma\rangle^{(\alpha)}_c$ and $D_\sigma^{(\alpha)}$ (\ref{D}). These two
quantities characterize size-distribution of the particles and distribution
function width, respectively.
In Figure 4 we present $\langle\sigma\rangle^{(\alpha)}_c$
and $D_\sigma^{(\alpha)}$ along the shadow curve.
It can be seen that the larger size
particles prefer the lower density phase and the smaller size particles are
predominantly encountered in the higher density phase. As the pressure increases
we observe increase in the mean size of the particles and decrease in the
distribution function width in the low-density phase and
decrease in the mean size of the particles and increase in the
distribution function width in the high-density
phase. The magnitude of these changes for the mean size of the particles in both
phases are similar (Fig 4a): we observe a strong sharpening of the low-density
phase distribution function with increasing pressure, while increase in the width
of the high-density phase distribution function do not exceed $13\%$ of its parent
phase value (Fig. 4b). A more detailed and systematic investigation  of these
effects for the mixture with polydispersity in both colloidal and polymeric
subsystems will be postponed to a future contributions.

\section{Conclusions}

With the concept presented above we are able to calculate the full phase diagram
of polydisperse hard-sphere colloidal and flexible chain particles with all types
of polydispersity.
For a particular system with polydispersity in the size of the colloidal particles
we present and discuss the phase diagram and corresponding fractionation effects.

\acknowledgments

One of the authors (YVK) was partially supported by the Science \& Technology
Center in Ukraine (project No. 4140). YVK gratefully acknowledges
the hospitality at the Vanderbilt University where part of this work was performed.


\begin{thebibliography}{99}

\bibitem{AO1}
\Name{Asakura S. \and Oosawa F.}
\Review{J. Chem. Phys.}
\Vol{22}
\Year{1954}
\Page{1255;}

\bibitem{AO2}
\Name{Asakura S. \and Oosawa F.}
\Review{J. Polym. Sci.}
\Vol{33}
\Year{1958}
\Page{183};

\bibitem{vrij}
\Name{Vrij A.}
\Review{Pure Appl. Chem.}
\Vol{48}
\Year{1976}
\Page{471;}

\bibitem{tuiner}
\Name{Tuinier R. \and Rieger J. \and de Kruif C.G.}
\Review{Adv. Colloid. Interface Sci.}
\Vol{103}
\Year{2003}
\Page{1;}

\bibitem{fuchs}
\Name{Fuchs M. \and Schweizer K.S.}
\Review{J. Phys. (Condens. Matt.)}
\Vol{14}
\Year{2002}
\Page{R239;}

\bibitem{paricaud1}
\Name{Paricaud P. \and Varga S. \and Jackson G.}
\Review{J. Chem. Phys.}
\Vol{118}
\Year{2003}
\Page{8525;}

\bibitem{zukowski}
\Name{Ramakrishnan S. \and Fuchs M. \and Schweizer K.S. \and Zukoski C.F.}
\Review{J. Chem. Phys.}
\Vol{116}
\Year{2002}
\Page{2201;}

\bibitem{paricaud2}
\Name{Paricaud P. \and Varga S. \and Cummings P.T. \and Jackson G.}
\Review{Chem. Phys. Lett.}
\Vol{398}
\Year{2004}
\Page{489;}

\bibitem{xu}
\Name{Bellier-Castella L. \and Xu H. \and Baus M.}
\Review{J. Chem. Phys.}
\Vol{113}
\Year{2000}
\Page{8337;}

\bibitem{kal1}
\Name{Kalyuzhnyi Yu.V. \and Kahl G.}
\Review{J. Chem. Phys.}
\Vol{119}
\Year{2003}
\Page{7335;}

\bibitem{wtd}
\Name{Wertheim M.S.}
\Review{J. Chem. Phys.}
\Vol{87}
\Year{1987}
\Page{7323}

\bibitem{gubbins}
\Name{Chapman W.G. \and Jackson G. \and Gubbins K.E.}
\Review{Mol. Phys.}
\Vol{65}
\Year{1988}
\Page{1057;}

\bibitem{boublik}
\Name{Boublik T.}
\Review{J. Chem. Phys.}
\Vol{53}
\Year{1970}
\Page{471;}

\bibitem{mansoori}
\Name{Mansoori G.A. \and Carnahan N.F. \and Starling K.E. \and Leland I.W.}
\Review{J. Chem. Phys.}
\Vol{54}
\Year{1971}
\Page{1523;}

\bibitem{sollich}
\Name{Sollich P.}
\Review{J. Phys. (Condens. Matt.)}
\Vol{14}
\Year{2002}
\Page{R79;}

\end{thebibliography}
\end{document}